\documentstyle[twoside,fleqn,espcrc2]{article}

\newcommand{\AmS}{{\protect\the\textfont2
  A\kern-.1667em\lower.5ex\hbox{M}\kern-.125emS}}

\def\Ds{D \!\!\!\! / \,}

\title{Coupling the Deconfining and Chiral Transitions}

\author{Peter~N.~Meisinger and Michael~C.~Ogilvie
\address{Department of Physics, Washington University, St. Louis, MO 63130}
\thanks{
This work was supported by the U.S. Department of Energy under grant number
DE-FG02-91-ER40628.}}

\begin{document}

\begin{abstract}
The Polyakov loop and the chiral condensate are used as order parameters to
explore analytically the possible phase structure of finite temperature QCD.
Nambu-Jona-Lasinio models in a background temporal gauge field are combined
with a Polyakov loop potential in a form suitable for both the lattice and the
continuum. Three possible behaviors are found: a first-order transition, a
second-order transition, and a region with both transitions.
\end{abstract}

\maketitle

\section{INTRODUCTION}
\label{s1}
Our recent work\cite{MeOg} on an effective action for dynamical quarks
has caused us to examine the simple picture in which deconfinement and chiral
symmetry are independent. On the basis of this work, it appears necessary to
consider models of the deconfining and chiral symmetry transitions in which the
two order parameters, the chiral condensate and the Polyakov loop, are coupled. 

In this paper we study a two flavor Nambu-Jona-Lasinio model \cite{NaJo} in a
uniform background temporal gauge field with a Euclidean action of the form
\begin{eqnarray}
{\cal L} = \bar{\Psi}(\Ds + m) \Psi
+ G \left[ (\bar{\Psi} \Psi )^2 + (\bar{\Psi} i \gamma_5 \vec{\tau} \Psi )^2
\right]
\label{e1.1}
\end{eqnarray}
where
\begin{eqnarray}
D_{\mu} = \partial_{\mu} + igA_{\mu}(x) \quad {\rm with} \quad
A_{\mu}(x) = \delta_{0\mu}A_0.
\label{e1.1a}
\end{eqnarray}
The spontaneous breakdown of $Z_3$ symmetry induced by the gauge field is
modeled by a simple potential with cubic and quartic terms. Combined with the
quark interaction effects, this model explicitly constructs a free energy
dependent on the two order parameters, in the spirit of Landau-Ginsberg
arguments. Because this model generates an effective potential consistent with
two-flavor QCD, we will show that the class of critical behaviors possible in
finite temperature QCD is {\it a priori} larger than has been previously
realized.

\section{EFFECTIVE POTENTIAL FOR FINITE TEMPERATURE NJL MODEL}
\label{s2}
We let $S_0$ denote the propagator for a free fermion with current mass $m$ and
$S$ denote the propagator for a fermion with constituent mass $M$.  Following
Cornwall, Jackiw, and Tomboulis \cite{CoJaTo}, a one-loop effective action
$\Gamma(S)$ for the theory given in Eq.~(\ref{e1.1}) at finite temperature is
given by
\begin{eqnarray}
\Gamma(S) = {\rm Tr} \left[ \ln \left( S_0^{-1} S \right) \right]
- {\rm Tr} \left( S_0^{-1} S  - 1 \right) \nonumber
\end{eqnarray}
\begin{eqnarray}
\quad \quad \quad \quad
- \tilde{G} \int_0^\beta dt \int d^3 \vec{x} \left[ {\rm Tr} S(x,x) \right]^2 
\label{e2.3}
\end{eqnarray}
where $\tilde{G} = [ 1 + 1 / (2 N_f N_c) ]G$. The effective potential can be
conveniently written in terms of the constituent mass M:
\begin{eqnarray}
V(M,T) = { (M - m)^2 \over 4 \tilde{G} }
\nonumber
\end{eqnarray}
\begin{eqnarray}
- {2 N_f \over \beta} \sum_{k_0} \int {d^3 \vec{k} \over (2 \pi)^3}
{\rm tr_c} \ln \Bigl[ (k_0 + gA_0)^2 + \omega_{\vec{k}}^2 \Bigr]
\label{e2.12}
\end{eqnarray}
where $\omega_{\vec{k}} = \sqrt{\vec{k}^2 + M^2}$.
Evaluating the mode sum over $k_0$, we find that up to an irrelevant constant
\begin{eqnarray}
V(M,T) = { (M - m)^2 \over 4 \tilde{G} }
-  2 N_f N_c \int {d^3 \vec{k} \over (2 \pi)^3} \omega_{\vec{k}} \nonumber
\end{eqnarray}
\begin{eqnarray}
+ {2 N_f \over \beta} \int {d^3 \vec{k} \over (2 \pi)^3}
{\rm tr_c} \ln \left( 1 + e^{- \beta \omega_{\vec{k}}} {\cal P} \right)
\nonumber 
\end{eqnarray}
\begin{eqnarray}
+ {2 N_f \over \beta} \int {d^3 \vec{k} \over (2 \pi)^3} {\rm tr_c}
\ln \left( 1 + e^{- \beta \omega_{\vec{k}}} {\cal P}^{\dagger} \right),
\label{e2.16}
\end{eqnarray}
where ${\cal P} = e^{i \beta g A_0}$ is the Polyakov loop associated with the
background $A_0$.

We regulate $V(M)$ by introducing a non-covariant cut-off, $\theta(\Lambda^2 -
\vec{k}^2)$, and make the approximation
\begin{eqnarray}
{\rm tr_c}({\cal P}^n) \approx N_c \left( {{\rm tr_c} {\cal P} \over N_c}
\right)^n.
\label{e2.19}
\end{eqnarray}
Noting that the quark determinant forces the trace of the Polyakov loop to be
real, Eq.~(\ref{e2.16}) now becomes
\begin{eqnarray}
V(M,T) = { (M - m)^2 \over 4 \tilde{G} }
- {N_f N_c \over \pi^2} \int_0^{\Lambda} dk \, k^2 \omega_k \nonumber
\end{eqnarray}
\begin{eqnarray}
+ {2 N_f N_c \over \pi^2 \beta} \int_0^{\Lambda} dk \, k^2
\ln \Bigl[ 1 + e^{- \beta \omega_k}
\left( {{\rm tr_c} {\cal P} \over N_c} \right) \Bigr].
\label{e2.20}
\end{eqnarray}
We determine $\Lambda$, $M$, and $\tilde{G}$ by simultaneously fixing the
chiral condensate and the pion form factor \cite{Be}, with $<\bar{\Psi}\Psi> =
N_f (246.7 MeV)^3$ and $f_{\pi} = 93 MeV$. For the remainder of this paper we
will use the numerical solution
\cite{Be,BeHaMe}
\begin{eqnarray}
M = 335.1 MeV \qquad \Lambda = 630.9 MeV \nonumber
\end{eqnarray}
\begin{eqnarray}
\tilde{G}\Lambda^2 = 2.196,
\label{e2.24}
\end{eqnarray}
which assumes a current quark mass $m_u = m_d = 4 MeV$.

\section{POLYAKOV LOOP POTENTIAL}
\label{s3}
We denote $tr_c {\cal P}$ by $\phi$. Near the deconfinement temperature $T_D$
the potential of Polyakov loops in pure QCD can be approximated by a polynomial
potential of the form \cite{TrWi}
\begin{eqnarray}
U(\phi ,T) = \lambda \left( \phi - {\phi^2 \over \phi_D} \right)^2 -
{T - T_D \over T_D L^{-1}} \left( {\phi \over \phi_D} \right)^3
\label{e3.1}
\end{eqnarray}
The potential is parametrized such that for a pure gauge theory the phase
transition occurs at $T_D$ with a latent heat $L$ and the Polyakov loop jumping
from $0$ to $\phi_D$.

This choice for $U$ is phenomenological. Although perturbative QCD does yield a
quartic polynomial potential for the $A_0$ field, the potential so obtained
does not display critical behavior, and it is only valid at high temperatures
\cite{We}.
However, it does suggest that the parameter $\lambda$ should be taken to be on
the order of $T_D^4$. We have chosen $\lambda = T_D^4$ and $L = 2T_D^4$ for the
numerical calculations below, neglecting any possible temperature corrections
or other dependencies in $\lambda$, $T_D$, $\phi_D$ and $L$. The direct
association of $T_D$, $\phi_D$ and $L$ with measureable quantities holds only
for very heavy quarks.  For light quarks, these parameters must be determined
by fitting to the observed behavior.

We can now couple the chiral symmetry and deconfinement phase transitions. Let
\begin{eqnarray}
{\cal V}(M,\phi,T) = V(M,\phi,T) + U(\phi,T),
\label{e3.3}
\end{eqnarray}
where $V(M,\phi,T)$ is given in Eq.~(\ref{e2.20}) with ${\rm tr_c}{\cal P} =
\phi$. To determine the critical behavior of the coupled effective potential,
the absolute minimum of the potential ${\cal V}(M,\phi,T)$ as a function of $M$
and $\phi$ must be found as $T$ is varied. A satisfactory determination of the
entire phase diagram requires numerical investigation.

\section{RESULTS}
\label{s4}
The two flavor Nambu-Jona-Lasinio model has a second-order chiral phase
transition for massless quarks. For the parameter set used here, this
transition occurs at $T_{\chi} = 194.6$ MeV, as determined numerically from
$V(M,\phi \equiv 1,T)$. The order of the transition is consistent with Monte
Carlo simulations of two-flavor QCD, and the critical temperature is plausible.
However, the finite temperature quark determinant is not consistent with the
known behavior of QCD unless the effects of a non-trivial Polyakov loop are
included. As Eq.~\ref{e2.16} makes clear, the effects of finite temperature in
the quark determinant are suppressed by the small expected value of the
Polyakov loop at low temperatures. Without the Polyakov loop effects, the
conventional Nambu-Jona-Lasinio model displays the $T^4$ behavior of a free
quark gas at arbitrarily low temperatures.

\begin{figure}[htb]
\bigskip
\bigskip
\bigskip
\bigskip
\bigskip
\bigskip
\bigskip
\bigskip
\bigskip
\bigskip
\bigskip
\bigskip
\bigskip
\bigskip
\bigskip
\bigskip
\bigskip
\bigskip
\bigskip
\bigskip
\bigskip
\includegraphics{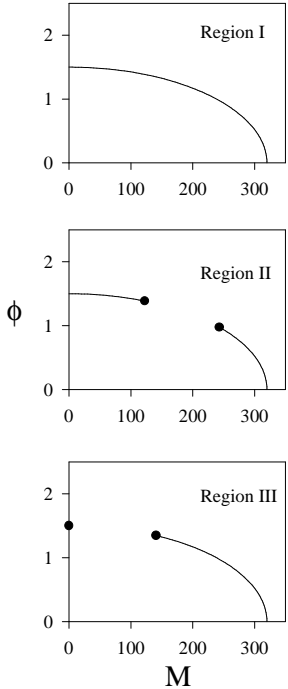}
\caption {Schematic behavior of $\phi$ and $M$ in the three regions of the
$(T_D, \phi_D)$ plane as temperature increases.}
\label{f1}
\end{figure}

Numerical investigation shows that the critical behavior of
${\cal V}(M,\phi,T)$ is most sensitive to $T_D$ and $\phi_D$. Figure~\ref{f1}
illustrates the three distinct types of critical behavior observed as $T_D$ and
$\phi_D$ are varied with the other parameters held fixed and the current mass
$m$ set to $0$. In Region I of the $(T_D, \phi_D)$ plane the effective quark
mass changes continuously from its $T = 0$ value to a mass of $0$ MeV as the
temperature increases, signaling a second-order chiral phase transition.
Simultaneously, the expectation value of the trace of the Polyakov loop
increases smoothly with temperature, exhibiting a large, but continuous,
increase in the vicinity of the chiral transition. This region exhibits
behavior most similar to that observed in simulations of two-flavor QCD. In
Region II the Polyakov loop experiences a first-order jump at some $T_c$. At
the same $T_c$ the constituent quark mass undergoes a sudden drop, but chiral
symmetry is {\it not} restored. As the temperature increases beyond $T_c$, the
constituent mass moves continuously to zero. Thus, there are two distinct phase
transitions in Region II, a first-order transition driven by the dynamics of
deconfinement, and a later second-order chiral symmetry restoring transition.
In Region III the Polyakov loop experiences a first-order jump at some $T_c$,
and simultaneously the effective quark mass drops suddenly to $0$ MeV,
indicating a single first-order transition which restores chiral symmetry.

\begin{figure}[htb]
\bigskip
\bigskip
\bigskip
\bigskip
\bigskip
\bigskip
\bigskip
\bigskip
\bigskip
\bigskip
\bigskip
\includegraphics{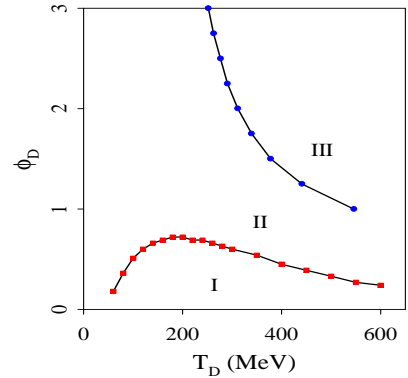}
\caption{Phase diagram in the $(\phi_D, T_D)$ plane.}
\label{f2}
\end{figure}

\begin{figure}[htb]
\bigskip
\bigskip
\bigskip
\bigskip
\bigskip
\bigskip
\bigskip
\bigskip
\bigskip
\includegraphics{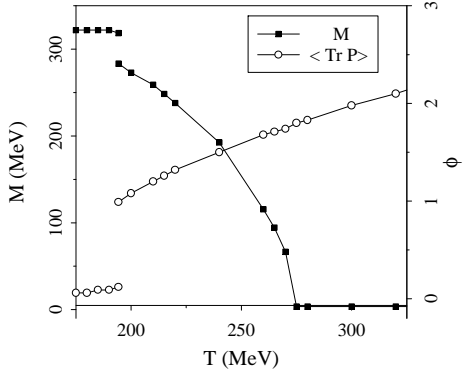}
\caption{Plot of $\phi$ and $M$ as a function of temperature at the point
$(T_D = 220 {\rm MeV}, \phi_D = 1.2)$ in Region II.}
\label{f3}
\end{figure}

Figure~\ref{f2} shows the location of the three regions in the $(T_D, \phi_D)$
plane. Figure~\ref{f3} plots the behavior of the Polyakov loop, $\phi$, and the
constituent quark mass, $M$, as a function of temperature at the point
$(T_D = 220 {\rm MeV}, \phi_D = 1.2)$ in Region II. There is a clear separation
of the first and second-order transitions.

\section{LATTICE VERSION OF THE EFFECTIVE POTENTIAL}
\label{s5}
A lattice version of the effective potential has the advantage that many
parameters can be set from QCD simulations. There are two related issues in
extending our work to the lattice: the choice of a lattice fermion formalism
and setting other parameters in the effective action.

A lattice version of the NJL model can be straightforwardly implemented using
naive fermions \cite{BiVr}. Species doubling can be handled by a formal
replacement of $N_c$ with $N_c / 16$. Using the single quark current mass,
chiral condensate, and constituent mass of the last section as inputs, we find
\begin{eqnarray}
a^{-1} = 389.0 MeV \qquad \tilde{G}a^{-2} = 0.8343.
\label{e5.8}
\end{eqnarray}
Hence, $f_{\pi} = 94.09$ $MeV$. Also note that $M_{T=0}a$ is of order 1.
The temperature is given by $T = 1 / (N_t a)$ where $N_t$ is the temporal
extent of the lattice. This places a problematic upper limit on a discrete
range of available temperatures. Given our choice of parameters,
$T_{max}$ is only $194.5$ $MeV$.

The temperature may be changed continuously by an asymmetric rescaling of the
lattice in which the lattice spacing $a_t$ in the temporal direction varies
independently of the spatial lattice spacing $a_s$ \cite{Ka}. Defining
$a_s = a$ and $\xi = a_t / a_s$, we now have in physical units
$T = 1 / (N_t a \xi)$. The effective potential is
\begin{eqnarray}
V(M, \phi, \xi) = \nonumber
\end{eqnarray}
\begin{eqnarray}
- {N_f N_c \over 4 \xi} \int_{-\pi}^{\pi} {d^3 \vec{k} \over (2 \pi)^3}
\rm{ln} \left[ \sqrt{1 + {\xi}^2 A^2(\vec{k})} + \xi A(\vec{k})
\right] \nonumber
\end{eqnarray}
\begin{eqnarray}
- {N_f N_c \over 2 N_t \xi} \int_{-\pi}^{\pi} {d^3 \vec{k} \over (2 \pi)^3}
\rm{ln} \Bigl\{ 1 + \Bigl[  \sqrt{1 + {\xi}^2 A^2(\vec{k})} \nonumber
\end{eqnarray}
\begin{eqnarray}
- \xi A(\vec{k}) \Bigr]^{N_t} {tr_c(\phi) \over N_c} \Bigr\}
+ {(M - m)^2 \over 4 \tilde{G}}
\label{e5.11}
\end{eqnarray}
where $A(\vec{k}) = \sqrt{M^2 + \sum_j {\sin}^2(p_j)}$.
However, using the same inputs as the symmetric lattice case, as well as
$f_{\pi} = 93$ $MeV$, we find the unique answer
\begin{eqnarray}
a^{-1} = 477.2 MeV \qquad \tilde{G}a^{-2} = 1.255
\label{e5.15}
\end{eqnarray}
with $\xi = 2.002$.

An alternative to asymmetric lattices might be variant or improved actions for
which $a^{-1}$ is larger.

\section{CONCLUSION}
\label{s6}
Polyakov loop effects have a strong impact on the possible critical behavior of
the NJL model. The universality argument \cite{PiWi} which predicts a
second-order chiral transition for two flavors and a first-order transition for
three or more flavors may fail when this additional order parameter is
included.

\end{document}